\documentclass[twocolumn,prb,nopacs,superscriptaddress]{revtex4}

\draft 

\usepackage{graphicx}
\usepackage{dcolumn}
\usepackage{bm}
\usepackage{times,mathptmx}
\usepackage{amsmath}
\usepackage{mhchem}
\usepackage{balance}

\begin{document}

\title{Substrate-free layer-number identification of two-dimensional materials: A case of Mo$_{0.5}$W$_{0.5}$S$_2$ alloy}


\author{Xiao-Fen Qiao}
\author{Xiao-Li Li}
\author{Xin Zhang}
\author{Wei Shi}
\author{Jiang-Bin Wu}
\author{Tao Chen}
\author{Ping-Heng Tan}
\email{phtan@semi.ac.cn}
\affiliation{State Key Laboratory of Superlattices and Microstructures, Institute of Semiconductors, Chinese Academy of Sciences, Beijing 100083, China}

\date{\today}

\begin{abstract}
Any of two or more two-dimensional (2D) materials with similar properties can be alloyed into a new layered material, namely, 2D alloy. Individual monolayer in 2D alloys are kept together by Van der Waals interactions. The property of multilayer alloys is a function of their layer number. Here, we studied the shear (C) and layer-breathing (LB) modes of Mo$_{0.5}$W$_{0.5}$S$_2$ alloy flakes and their link to the layer number of alloy flakes. The study reveals that the disorder effect is absent in the C and LB modes of 2D alloys, and the monatomic chain model can be used to estimate the frequencies of the C and LB modes. We demonstrated how to use the C and LB mode frequency to identify the layer number of alloy flakes deposited on different substrates. This technique is independent of the substrate, stoichiometry, monolayer thickness and complex refractive index of 2D materials, offering a robust and substrate-free approach for layer-number identification of 2D materials.
\end{abstract}


\maketitle 


Monolayer (SLG) and multilayer graphenes (MLGs) have attracted much attention due to its extraordinary properties\cite{Novoselov-Science-2004,Geim-nm-2007} and potentials in device applications\cite{bonaccorso-NP-2010-graphene}. Besides SLG and MLGs, there are a large number of other two-dimensional (2D) systems with interesting properties. For example, NiTe$_2$ and VSe$_2$ are semimetals; WS$_2$, WSe$_2$, MoS$_2$, MoSe$_2$ and MoTe$_2$ are semiconductors; h-BN, and HfS$_2$ are insulators; NbS$_2$, NbSe$_2$, and TaSe$_2$ are superconductors; Bi$_2$Se$_3$ and Bi$_2$Te$_3$ are topological insulators.\cite{Xu-CR-2013,Geim-nature-2013} Two or more layered materials with similar properties can be alloyed into a new layered material, such as Mo$_x$W$_{1-x}$S$_2$, Mo$_{1-x}$W$_x$Se$_2$ and (Bi$_{1-x}$Sb$_x$)$_2$Te$_3$ alloys.\cite{Xue-nc-11,Komsa-jpcl-12,Tongay-apl-14,chenyf-acsn-2013} The 2D alloys are a rich source of 2D materials because they can exhibit tunable properties because the ratio of two end compositions can be controllable.\cite{Xue-nc-11,chenyf-acsn-2013} The key to form a good quality of 2D monolayer is mixing the end compositions at atomic scale. Individual monolayers in the 2D alloys are kept together by Van der Waals interations. Similar to graphite and graphene, the property of multilayer alloys is a function of their layer number. How to determine the layer number of ultrathin 2D alloys is thus of primary importance for fundamental science and applications.

Several optical techniques have been developed to identify the layer number of 2D materials.\cite{Nizh-nl-2007,Casiraghi-nl-2007,Blake-apl-2007,Yoon-prb-2009,Lee-acsnano-2010,mak-prl-2010,XL-NS-2015} Nevertheless, most techniques rely on the specific properties of 2D materials. For example, one can identify the layer number, denoted as N, of MoS$_2$  by the intensity or peak positions of the corresponding photoluminescence (PL) peak and Raman modes.\cite{Lee-acsnano-2010,mak-prl-2010} The mostly used technique is based upon optical contrast of 2D materials against dielectric substrates, such as a Si substrate covered with a SiO$_2$ layers.\cite{Blake-apl-2007,Nizh-nl-2007} To precisely identify N of few-layer 2D materials, the experimental optical contrast must be compared with the theoretical one for different N. However, the theoretical calculations require pre-determined parameters, including  thickness of a single layer, complex refractive index of the material and information of the substrate,\cite{Casiraghi-nl-2007,hanwp-APS-2013} which poses great challenge for research on new 2D materials in early stage. Therefore, it is desirable to develop a novel approach to identify the layer number of few-layer 2D materials, relying on as little prerequisite parameters as possible.

The interlayer modes of multilayer 2D materials originates from the relative motions of the rigid monolayer planes themselves, either perpendicular or parallel to their normal, such as the shear (C) modes and the layer breathing (LB) modes.\cite{tanph-nm-2012,Plechinger-apl-12,zhangx-prb-2013,Zhaoyy-nl-2013,lui-nl-2014,wujb-natcom-2014,Zhang-csr-15}  The C and LB modes are the collective vibration modes of all the layers so their frequency ($\omega_C$ and $\omega_{LB}$) depends on N and the interlayer coupling.\cite{tanph-nm-2012,zhangx-prb-2013} By probing such ultralow-frequency (ULF) modes in multilayer 2D materials, it is possible to identify its layer number no matter what kinds of substrates is used to support multilayer flakes. In common alloys, random distribution of atoms will result in disorder effect on their vibration modes in terms of peak weakening and broadening.\cite{chenyf-ns-2014} For multilayer 2D alloys exfoliated from the bulk layered alloys, how the alloy layer modifies their C and LB modes is an open issue. Whether the monatomic chain model (MCM) based on 2D crystals\cite{zhangx-prb-2013} can be applied to the C and LB modes in multilayer alloys for layer-number identification is not clear.

Here, Mo$_{0.5}$W$_{0.5}$S$_2$ alloy has been used as a template, to explore the possibility of applying MCM to 2D alloys. Mo$_{1-x}$W$_x$S$_2$ alloy will be simplified as MoWS$_2$ alloy hereafter for simplicity. It is demonstrated that, by only measuring the C and LB modes, the layer number of MoWS$_2$ alloy flakes up to N=11 can be determined rapidly and nondestructively. Because the frequencies of the C and LB modes are linked with the layer number by a robust general formula based on MCM, this technique can be applied to ultrathin 2D materials on any substrate.

Ultra-thin flakes have been  mechanically exfoliated from bulk MoWS$_2$ alloy (purchased from 2d Semiconductors Inc.), and transferred onto a Si substrate covered with 90-nm SiO$_2$ film.\cite{Novoselov-Science-2004} Raman and PL measurements are performed at room temperature using a Jobin-Yvon HR800 micro-Raman system equipped with a liquid-nitrogen-cooled charge couple detector (CCD), a $\times$ 100 objective lens (numerical aperture=0.90) and a 1800 lines/mm grating. The excitation wavelength is 466 nm from a Ar$^+$ laser. At this wavelength, the 1800 lines/mm grating enables us to have each pixel of the charge-coupled detector cover 0.35 cm$^{-1}$. Plasma lines have been removed from the laser beam, using BragGrate Bandpass filters. Measurements down to 5 cm$^{-1}$ are enabled by three BragGrate notch filters with optical density 3 and with FWHM=5cm$^{-1}$. Both BragGrate bandpass and notch filters are produced by OptiGrate Corp. The typical laser power is about 0.4 mW, to avoid sample heating.

\begin{figure}[!htb]
\centerline{\includegraphics[width=85mm,clip]{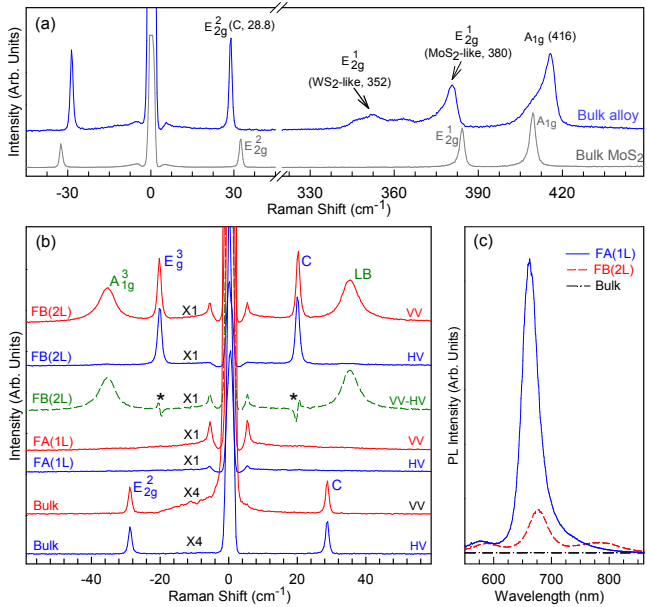}}
\caption{(a) Raman spectrum of bulk MoWS$_2$ alloy and bulk MoS$_2$ crystal in the ULF and 320-450cm$^{-1}$ ranges. (b) ULF Raman spectra of bulk, 1L and 2L alloy flakes under HV and VV polarization configurations, where the features labeled by stars are residual of the C mode after the spectrum subtraction. (c) PL spectra of bulk, 1L and 2L alloy flakes. The excitation wavelength is 466nm. } \label{Fig1}
\end{figure}

2H-MoWS$_2$ bulk alloys are formed by stacking monolayer (1L) alloy together via van der Waals interactions.\cite{chenyf-acsn-2013} The 1L alloy contains one MoW plane sandwiched by two S planes, represented as S-MoW-S. Flakes containing N S-MoW-S layer are denoted as NL MoWS$_2$. Raman spectrum of the bulk alloy is depicted in Fig. 1(a). Since, the point group of the bulk alloy is $D_{6h}$, the mode at 416cm$^{-1}$ is assigned as A$_{1g}$,\cite{chenyf-ns-2014}, as it lies between A$_{1g}$ of MoS$_2$ (409cm$^{-1}$) and WS$_2$ (420cm$^{-1}$). There are two modes at 380 cm$^{-1}$ and 352 cm$^{-1}$, assigned as MoS$_2$-like and WS$_2$-like E$^1_{2g}$,\cite{chenyf-ns-2014} respectively. The peaks at 347 and 365 cm$^{-1}$ may be the disorder-related Raman peaks.\cite{chenyf-ns-2014} The A$_{1g}$ and E$^1_{2g}$ peaks exhibit broadened profiles due to the disorder effect in the alloy compared with that (gray curves) in bulk MoS$_2$, as shown in Fig. 1(a). Only one ULF peak is observed at 28.8 cm$^{-1}$. Considering that the LB mode is Raman inactive and the C mode is Raman active in bulk MoS$_2$,\cite{Zhang-csr-15} this peak is attributed to the C mode in bulk alloys. Indeed, this peak is observable under both the parallel (VV) and cross (HV) polarization configuration, as shown in Fig. 1(b), similar to the case in bulk MoS$_2$. In contrast to the broadening of the high-frequency mode, full width at half maximum (FWHM) of the C mode in bulk alloys is about 1.1 cm$^{-1}$ and almost identical to that in bulk MoS$_2$, as displayed in Fig. 1(a). It  suggests that the disorder effect is absent in the C and LB modes of 2D alloy materials. Indeed, as the C mode is due to rigid-layer lattice vibrations, it is reasonable that the random distribution of Mo and W atoms in the MoW layer has little influence on the interlayer vibration modes.

The observation of the C mode in bulk alloys allows the deduction of $\omega_C$ of NL flakes, thus enabling the N determination of multilayer MoWS$_2$ alloys. Because the disorder effect is absent for the C mode in bulk MoWS$_2$ alloys, it is expected that it is also absent for the C and LB modes in NL alloys and the MCM can be applicable for NL alloys by treating each S-MoW-S layer as a single ball.\cite{zhangx-prb-2013} Therefore,the symmetry of NL MoWS$_2$ alloys can be analyzed in a similar way as of NL MoS$_2$ crystal.\cite{Zhang-csr-15} For the NL flake, there are N-1 C modes and N-1 LB modes. Based on the MCM, the N-dependent frequencies of both the C and LB modes exhibit as a fan diagrams.\cite{zhangx-prb-2013} There exist a series of upper and lower phonon branches originating from even N. The observed upper branches of the C and LB modes originating from 2L are denoted as $C_2^+$ and $LB_2^+$, respectively. Similar to the case of NL MoS$_2$,\cite{zhangx-prb-2013} the N-dependent frequencies of $C_2^+$ and $LB_2^+$ branches are, respectively, given by
\begin{equation}
\begin{split}
\omega(C_2^+)(N)=\omega_{C}(bulk)\cos(\pi/2N),\\
\omega(LB_2^+)(N)=\omega_{LB}(bulk)\cos(\pi/2N),
\end{split}
\label{Eq1}
\end{equation}
\noindent where $(N\ge2)$, $\omega_{C}(bulk)$ and $\omega_{LB}(bulk)$ are the C and LB frequency of bulk alloy, respectively. $\omega_C(bulk)=\sqrt2\omega_{C}(2L)$, and $\omega_{LB}(bulk)=\sqrt2\omega_{LB}(2L)$. Both $\omega(C_2^+)$(N) and $\omega(LB_2^+)$(N) stiffen with increasing N.\cite{zhangx-prb-2013}

At first, we will demonstrate how to identify 1L and 2L alloys from the observed $\omega_C(bulk)$. According to the $D_{3d}$ symmetry for 2L MoWS$_2$, both the C and LB modes ,  denoted as E$^3_{g}$ and A$^3_{1g}$, respectively. We measured the ULF modes of many ultrathin alloy flakes with low optical contrast. There exist two typical sets (FA and FB) of ultrathin alloy flakes. Each set exhibits an identical Raman spectral feature, as depicted in Fig. 1(b). It is obvious that the FA flake is 1L alloy because no any peak is present above 5$cm^{-1}$ in its ULF range under VV and HV configurations.\cite{tanph-nm-2012} Under the VV configuration, the FB flake exhibits two ULF peaks located at 20.2 cm$^{-1}$ and 35.4 cm$^{-1}$, respectively, while only the peak at 20.2 cm$^{-1}$ appears under the HV configuration, and equals to exactly $1/\sqrt2$ times as much as $\omega_C(bulk)$ at 28.8 cm$^{-1}$ as given by Eq. (\ref{Eq1}). Thus, it is indicated that the FB flake is 2L alloy, and the 20.2 cm$^{-1}$ peak corresponds to its C mode. The other peak is then ascribed to LB mode with $\omega_{LB}(2L)$=35.4 cm$^{-1}$. Indeed, the calculated C and LB mode frequencies of bulk MoWS$_2$ alloys based on the MCM\cite{zhangx-prb-2013} are 20.2 and 35.6 cm$^{-1}$ respectively, if the rigid S-MoW-W layer takes the average mass of $(m_{Mo}+m_{W})/2+2m_{S}$ and interlayer force constants\cite{zhangx-prb-2013} of $\alpha^\parallel_{SS}$ and $\alpha^\perp_{SS}$ in MoS$_2$, where $m_{Mo}$, $m_{W}$ and $m_{S}$ are the masses of Mo layer in MoS$_2$, W layer in WS$_2$ and S layer in MoS$_2$, respectively. The intrinsic LB mode in 2L alloy can be obtained by subtracting the spectrum under the HV configuration from that under the VV configuration (denoted as VV-HV), as shown by the dashed curve in Fig. 1(b). Based on Eq. (\ref{Eq1}), we have $\omega_{LB}(bulk)$=50.1 cm$^{-1}$. To further support the layer number identification, the photoluminescence (PL) spectra of FA and FB have been measured., as demonstrated in Fig. 1(c). The FA flake exhibits a strong PL emission at 663 nm and no sideband appears at its lower-energy range. However, the FB flake exhibits a weaker PL emission at 678 nm and there is a low-energy sideband at 768 nm, which is assigned to the indirect PL emission of the 2L alloy. PL emission is not observed in bulk alloys at all. All the PL features are analogous to those in 1L, 2L and bulk MoS$_2$.\cite{mak-prl-2010}

\begin{figure}[!htb]
\centerline{\includegraphics[width=85mm,clip]{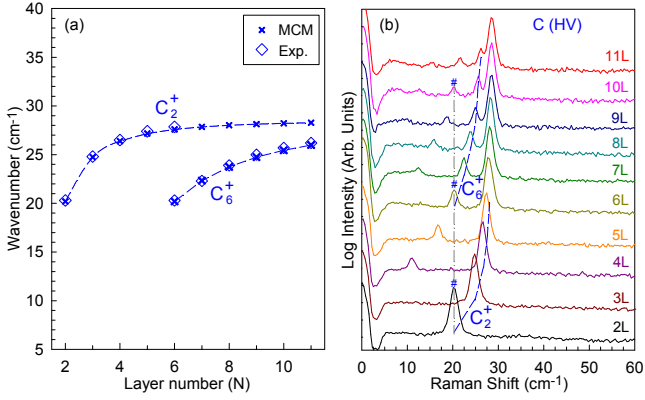}}
\caption{(a) Theoretical (crosses) and experimental (diamonds) frequencies of $C_2^+$ and $C_6^+$ phonon branches of 2L-11L alloys. The three \# indicate the C modes in 6L and 10L alloys with frequencies of $\omega_{C}(2L)$, as indicated by the dash-dotted lines. The dashed lines are guides to the eye.}  \label{Fig2}
\end{figure}

The above discussion reveals that one do apply MCM for the C mode in 2L alloys, thus we can apply MCM to the C and LB modes of NL alloys without doubt. The predicted frequencies of the $C_{2}^+$ branch based on Eq. (\ref{Eq1}) are plotted in Fig. 2(a). It shows that when N$>$6, the dependency of $\omega(C_2^+)(N)$ on N weakens with increasing N, so the difference between $\omega(C_2^+)(N)$ and $\omega(C_2^+)(N-1)$ are too close to be distinguishable. In this case, we can consider another upper phonon branch of the C modes originating from the 6L alloy, which is denoted as $C_6^+$. The N-dependent frequencies of the $C_6^+$ branch is given by\cite{zhangx-prb-2013}
\begin{equation}
\omega(C_6^+)(N)=\omega_{C}(bulk)\cos(3\pi/2N),
\label{Eq2}
\end{equation}
\noindent where $(N\ge6)$. $\omega(C_6^+)$(N) stiffens from $\omega_{C}(2L)$ with increasing N.\cite{zhangx-prb-2013} The predicted frequencies of the $C_6^+$ branch are also summarized in Fig. 2(a) by crosses. By comparing the experimental frequencies of the two branches $C_{2}^+$ and $C_{6}^+$ with theoretical ones, 2L-11L alloys can be easily identified. Detailed Raman spectra related with the C modes of the identified 2L-11L alloys are depicted in Fig. 2(b). The experimental $\omega(C_2^+)(N)$ (2$\leq$N$\leq$6) and $\omega(C_6^+)(N)$ (6$\leq$N$\leq$11) are summarized in Fig. 2(a) by diamonds, matching theoretical results very well.

Besides the C modes, there are a series of LB modes in NL alloys analogous to the case in NL-MoS$_2$.\cite{zhangx-prb-2013} Because the LB mode in bulk alloy is Raman inactive, the upper phonon branches originating from 2L and 6L cannot be observed in MoS$_2$ and similar 2D materials.\cite{zhangx-prb-2013} To the contrary, the lower phonon branches originating from 2L and 6L can be easily observed, which are denoted as $LB_2^-$ and $LB_6^-$, respectively, whose N-dependent frequencies are given by\cite{zhangx-prb-2013}

\begin{equation}
\begin{split}
\omega(LB_2^-)(N)=\omega_{LB}(bulk)\sin(\pi/2N), (N\ge2),\\
\omega(LB_6^-)(N)=\omega_{LB}(bulk)\sin(3\pi/2N),(N\ge6).
\end{split}
\label{Eq3}
\end{equation}

\noindent Both $\omega(LB_2^-)$(N) and $\omega(LB_6^-)$(N) softens from $\omega_{LB}(2L)$ with increasing N.\cite{zhangx-prb-2013} The predicted frequencies of the two branches are summarized in Fig. 3(a). Similar to the case of C modes, one can also identify N for NL alloys based on the N-dependent $\omega(LB_2^-)(N)$ or $\omega(LB_6^-)(N)$. Raman spectra of the LB modes of the identified 2L-11L alloys are depicted in Fig. 3(b). The experimental $\omega(LB_2^-)(N)$ (2$\leq$N$\leq$11) and $\omega(LB_6^-)(N)$ (6$\leq$N$\leq$11) have been extracted and plotted in Fig. 3(a) for comparison with theoretical predictions. The large separation between $\omega(LB_2^-)(N)$ and $\omega(LB_2^-)(N+1)$ ensures that one can identify N of MoWS$_2$ alloy up to 11L only by the $LB_2^-$ branch. If $\omega_{LB}(2L)$ is too small in some 2D materials, $\omega(LB_2^-)(N)$ may quickly approach to 5-10 cm$^{-1}$ , the detection limit of a typical Raman system. Similar to the case of the C mode in MoWS$_2$ alloy, one must use the $LB_6^-$ branch to identify N of thicker 2D flakes.

\begin{figure}[!htb]
\centerline{\includegraphics[width=85mm,clip]{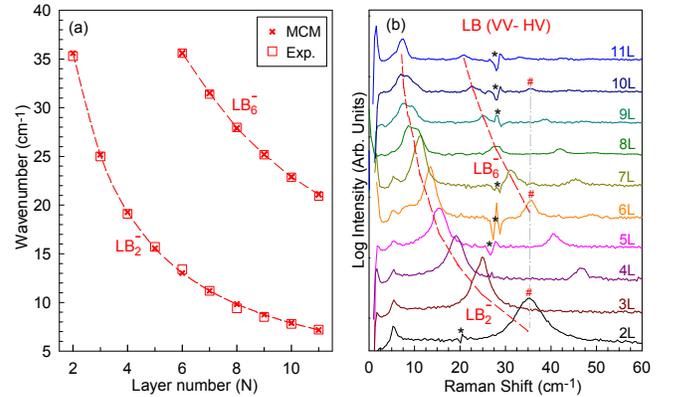}}
\caption{(a) Theoretical (crosses) and experimental (squares) frequencies of $LB_2^-$ and $LB_6^-$ phonon branches of 2L-11L alloys. The three \# indicate the LB modes in 6L and 10L alloys with frequencies of $\omega_{LB}(2L)$, as indicated by the dash-dotted lines. The features labeled by stars in (b) are residual signals of the C mode after spectrum subtraction. The dashed lines are guides to the eye.} \label{Fig3}
\end{figure}

We have demonstrated how to apply MCM for identifying layer-number of ultrathin alloy flakes. Although the Mo layer in multilayer MoS$_2$ was replaced by an alloy layer with random distribution of Mo and W atoms to form multilayer MoWS$_2$ alloy, the C and LB modes in multilayer alloys can be well-understood by MCM that describes the rigid interlayer vibrations in 2D crystals. This means that MCM is adequate for various 2D alloys that be combined by any two end materials. In principle, the above standard identification process can be used for common 2D crystals. Once the C$_2^+$ or LB$_2^-$ branch is detected, layer number of ultra-thin 2D materials can be identified. $\omega_{C}(bulk)$ of most attractive 2D materials is larger than 20 cm$^{-1}$ and $\omega_{LB}(bulk)$ is usually larger than $\omega_{C}(bulk)$. In this case, layer number of ultrathin 2D materials up to 5L can be identified from C$_2^+$ or LB$_2^-$ branch with a high-resolution Raman system. One can further identify layer number of thicker flakes by examining C$_6^+$ or LB$_6^-$ branch once they are detected.

\begin{figure}[!htb]
\centerline{\includegraphics[width=85mm,clip]{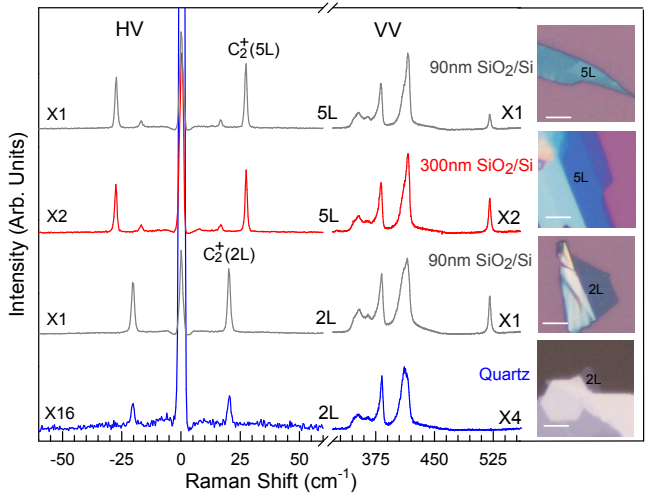}}
\caption{The C modes and high-frequency modes of MoWS$_2$ alloy flakes on quartz, 300nm SiO$_2$/Si, and 90nm SiO$_2$/Si, which were measured under HV and VV configurations, respectively. The inset shows the corresponding optical images. The scale bar in each image is $3{\mu}m$.} \label{Fig4}
\end{figure}

The C and LB vibrations are an intrinsic property of multilayer 2D materials, which are irrelevant to any substrate. Indeed, the substrate information is not involved in the above discussion at all. Therefore, this layer-number identification method proves to be substrate-free for 2D materials. As an example, Fig. 4 shows the C and high-frequency modes of four MoWS$_2$ alloy flakes exfoliated on three types of substrates including quartz, 300nm SiO$_2$/Si, and 90nm SiO$_2$/Si. Using the aforementioned method, the layer numbers of the corresponding flakes have been identified to be 2L and 5L. Their optical contrasts in their optical images are very different from each other because of different N and substrates,\cite{Casiraghi-nl-2007,hanwp-APS-2013} as shown in the inset to Fig. 4. The high-frequency modes of the four alloys exhibit similar spectral features as expected. However, the alloy flakes with the same layer number exhibit the C modes with almost the same frequency, in spite of the significant peak intensity due to the optical interference effect in the MoWS$_2$/SiO$_2$/Si multilayer.\cite{Yoon-prb-2009,XL-NS-2015} Alloy flakes with different layer number display disparate C modes, offering a reliable diagnostic of the layer number. Similar behavior has also been observed for the LB modes (not shown here). It is evident that the identification of the layer number is not dependent on the substrates. Moreover, Eqs.(\ref{Eq1})-(\ref{Eq3}) are not relevant to the stoichiometry, monolayer thickness and complex refractive index of any 2D materials. Therefore, this method can be applied to other 2D materials once $\omega_{C}(bulk)$ or $\omega_{LB}(bulk)$ is determined by Raman measurement.

In summary, we used MoWS$_2$ as a template to study the N-dependent ultralow-frequency Raman spectroscopy of 2D alloys. The interlayer C and LB modes can be observed in multilayer MoWS$_2$ alloys where the alloy effect can be ignored. $\omega_{C}$ and $\omega_{LB}$ in NL alloys are linked with N by robust formulas based on MCM. We have demonstrated how to identify the layer number of NL alloy by their C or LB modes once $\omega_{C}(bulk)$ or $\omega_{LB}(bulk)$ is determined by Raman measurement. In principle, this technique can be applied to other 2D materials produced by micro-mechanical exfoliations, chemical-vapor-deposition growth or transfer processes on various substrates, without the need for monolayer thickness and complex refractive index of 2D materials and the supporting substrate..

This work was supported by the National Natural Science Foundation of China, grants 11225421, 11434010 and 11474277.


\begin{thebibliography}{10}%
\bibitem{Novoselov-Science-2004} K. S. Novoselov, A. K. Geim, S. V. Morozov, D. Jiang, Y. Zhang, S. V. Dubonos, I. V. Grigorieva, and A. A. Firsov, Science {\bf306}, 666 (2004).
\bibitem{Geim-nm-2007} A. K. Geim and K. S. Novoselov, Nat. Mater. {\bf6}, 183 (2007).
\bibitem{bonaccorso-NP-2010-graphene} F.Bonaccorso, Z. Sun, T. Hasan, and A. Ferrari, Nat. Photon. {\bf4}, 611 (2010).
\bibitem{Xu-CR-2013} M. S. Xu, T. Liang, M. M. Shi, and H. Z. Chen, Chem. Rev. {\bf113}, 3766 (2013).
\bibitem{Geim-nature-2013} A. K. Geim and I. V. Grigorieva, Nature {\bf499}, 419 (2013).
\bibitem{Xue-nc-11} J. S. Zhang, C. Z. Chang, Z. C. Zhang, J. Wen, X. Feng, K. Li, M. H. Liu, K. He, L. L. Wang, X. Chen, and et al., Nat. Commun. {\bf2}, 574 (2011).
\bibitem{Komsa-jpcl-12} H. -P. KomsaandA. and V. Krasheninnikov, J. Phys. Chem. Lett. {\bf3},3652 (2012).
\bibitem{Tongay-apl-14} S. Tongay, D. S. Narang, J. Kang, W. Fan, C. Ko, A. V. Luce, K. X. Wang, J. Suh, K. D. Patel, V. M. Pathak, and et al., Appl. Phys. Lett. {\bf104}, 012101 (2014).
\bibitem{chenyf-acsn-2013} Y. F. Chen, J. Y. Xi, D. O. Dumcenco, Z. Liu, K. Suenaga, D. Wang, Z. G. Shuai, Y.-S. Huang, and L. M. Xie, ACS Nano {\bf7}, 4610 (2013).
\bibitem{Nizh-nl-2007} Z. H. Ni, H. M. Wang, J. Kasim, H. M. Fan, T. Yu, Y. H. Wu, Y. P. Feng, and Z. X. Shen, Nano Lett. {\bf7}, 2758 (2007).
\bibitem{Casiraghi-nl-2007} C. Casiraghi, A. Hartschuh, E. Lidorikis, H. Qian, H. Harutyunyan, T. Gokus, K. S. Novoselov, and A. C. Ferrari, Nano Lett. {\bf7}, 2711 (2007).
\bibitem{Blake-apl-2007} P. Blake, E. W. Hill, A. H. C. Neto, K. S. Novoselov, D. Jiang, R. Yang, T. J. Booth, and A. K. Geim, Appl. Phys. Lett. {\bf91}, 063124 (2007).
\bibitem{Yoon-prb-2009} D. Yoon, H. Moon, Y.-W. Son, J. Choi, B. Park, Y. Cha, Y. H. Kim, and H. Cheong, Phys. Rev. B {\bf80}, 125422 (2009).
\bibitem{Lee-acsnano-2010} C. Lee, H. Yan, L. E. Brus, T. F. Heinz, J. Hone, and S. Ryu, ACS Nano {\bf4}, 2695 (2010).
\bibitem{mak-prl-2010} K. F. Mak, C. Lee, J. Hone, J. Shan, and T. F. Heinz, Phys. Rev. Lett. {\bf105}, 136805 (2010).
\bibitem{XL-NS-2015} X. L. Li, X. F. Qiao, W.-P. Han, Y. Lu, Q.-H. Tan, X.-L. Liu, and P.-H. Tan, Nanoscale {\bf7}, 8135 (2015).
\bibitem{hanwp-APS-2013} W. P. Han, Y. M. Shi, X. L. Li, S. Q. Luo, Y. Lu, and P. -H. Tan, Acta Phys. Sin. {\bf62}, 110702 (2013).
\bibitem{tanph-nm-2012} P. -H. Tan, W. -P. Han, W. -J. Zhao, Z. -H. Wu, K. Chang, H. Wang, Y. F. Wang, N. Bonini, N. Marzari, N. Pugno, and et al., Nat. Mater. {\bf11}, 294 (2012).
\bibitem{Plechinger-apl-12} H. S. E. J. W. D. S. C. Plechinger, G. and T. Korn, Appl. Phys. Lett. {\bf101}, 101906 (2012).
\bibitem{zhangx-prb-2013} X. Zhang, W. P. Han, J. B. Wu, S. Milana, Y. Lu, Q. Q. Li, A. C. Ferrari,
and P. H. Tan, Phys. Rev. B {\bf87}, 115413 (2013).
\bibitem{Zhaoyy-nl-2013} Y. Y. Zhao, X. Luo, H. Li, J. Zhang, P. T. Araujo, C. Gan, J. Wu, H. Zhang, S. Quek, M. Dresselhaus, and Q. Xiong, Nano Lett. {\bf13}, 1007 (2013).
\bibitem{lui-nl-2014} C. H. Lui, Z. P. Ye, C. Keiser, X. Xiao, and R. He, Nano Lett. {\bf14}, 4615 (2014).
\bibitem{wujb-natcom-2014} J. B. Wu, X. Zhang, M. Ij$\ddot{a}$s, W. P. Han, X. F. Qiao, X. L. Li, D. S. Jiang, A. C. Ferrari, and P. H. Tan, Nat. Commun. {\bf5}, 5309 (2014).
\bibitem{Zhang-csr-15} X. Zhang, X.-F. Qiao, W. Shi, J.-B. Wu, D.-S. Jiang, and P.-H. Tan, Chem. Soc. Rev. {\bf44}, 2757(2015).
\bibitem{chenyf-ns-2014} Y. F. Chen, D. O. Dumcenco, Y. M. Zhu, X. Zhang, N. N. Mao, Q. L. Feng, M. Zhang, J. Zhang, P.-H. Tan, and et al., Nanoscale {\bf6}, 2833 (2014).

\end{thebibliography}
\end{document}